\documentclass[showpacs,amsmath,twocolumn,showkeys,pra]{revtex4-1}
\usepackage{dcolumn}    
\usepackage{bm}         
\usepackage{ifpdf}
\usepackage{amssymb,lineno,amsfonts}
\usepackage{graphicx}   
\usepackage{bbm}
\usepackage{mathrsfs}
\usepackage{upgreek}
\usepackage{epstopdf}
\usepackage{setspace}
\usepackage{hyperref}
\usepackage[matrix,frame,arrow]{xypic}
\usepackage{float}
\usepackage{natbib}
\usepackage{color} 
\usepackage{mathbbold}
\definecolor{med-blue}{RGB}{25,25,112} 
\hypersetup{colorlinks, linkcolor={red},citecolor={blue}, urlcolor={blue}}

\newcommand{\ket}[1]{\vert{#1}\rangle}

\begin{document}

\title{Steering Quantum Dynamics via Bang-Bang Control: \\
Implementing optimal fixed point quantum search algorithm}
\author{Gaurav Bhole, Anjusha V. S., and T. S. Mahesh
}
\email{mahesh.ts@iiserpune.ac.in}
\affiliation{Department of Physics and NMR Research Center,\\
Indian Institute of Science Education and Research, Pune 411008, India}

\begin{abstract}
{A robust control over quantum dynamics is of paramount importance for quantum technologies.  
	Many of the existing control techniques are based on smooth Hamiltonian modulations involving repeated calculations of basic unitaries resulting in time complexities scaling rapidly with the length of the control sequence. 	On the other hand, the bang-bang controls need one-time calculation of basic unitaries and hence scale much more efficiently.  By employing a global optimization routine such as the genetic algorithm, it is possible to synthesize not only highly intricate unitaries, but also certain nonunitary operations.
	Here we demonstrate the unitary control through the first implementation of the optimal fixed-point quantum search algorithm in a three-qubit NMR system.  More over, by combining the bang-bang pulses with the twirling process, we also demonstrate a nonunitary transformation of the thermal equilibrium state into an effective pure state in a five-qubit NMR system.
}
\end{abstract}

\maketitle

{\it Introduction:}
Quantum physics offers an exciting platform for the next generation devices.  While quantum computers promise a tremendous computational capability, other quantum devices have opened up a variety of interesting possibilities such as quantum imaging \cite{image,qimage}, supersensitive field sensors \cite{sensors,PhysRevLett.93.173002,jones2009magnetic}, and  the generation of random numbers \cite{jennewein2000fast,dynes2008high}.
Robust and efficient quantum control is indispensable not only for quantum computers \cite{divincenzo}, but also for other quantum simulators \cite{RevModPhys.86.153}.

Over the past several years a remarkable progress has been achieved in steering quantum dynamics 
(eg. \cite{SMP,GRAPE,mahesh2006quantum,tovsner2009optimal,wang2009entanglement}).  Typically one considers a constant internal Hamiltonian ${\cal H}_\mathrm{int}$ and a set of $M$ control operators $\{h_j\}$ such that the total Hamiltonian is of the form
${\cal H}(t) = {\cal H}_\mathrm{int} + \sum_{j=1}^M u_{j}(t) h_j$, where $\{u_j(t)\}$ form a set of 
time-dependent parameter profiles.
The overall unitary of duration $T$ is
$U = D \exp \left(-i \int_{0}^{T}{\cal H}(t) dt\right) $, where $D$ is the Dyson time ordering operator.
Given a target unitary $U_T$ of dimension $N$, the goal is to optimize 
$\{u_{j}(t)\}$, by maximizing the unitary fidelity $F_u = \vert \mathrm{Tr}(U_T^\dagger U)/N \vert^2$.

\begin{figure}[b]
	\includegraphics[trim=0cm 6.7cm 0cm 4.2cm, clip=true,width=9cm]{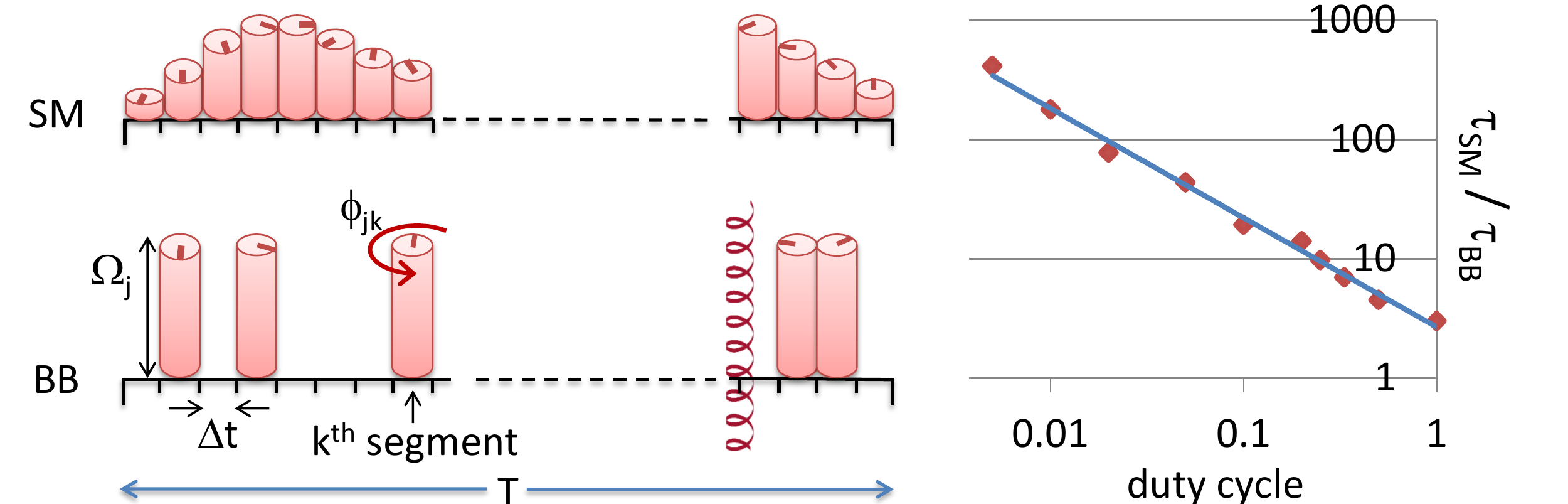} 
	\caption{Smooth modulation (SM) vs bang-bang (BB) sequence.  
		$\Omega_j$, $\phi_{jk}$, $\Delta t$ are the amplitude, phase, and duration of the segments,
		and $T$ is the total duration of the sequence.  The helix represents the twirling operation
		required for nonunitary gates.
		Performance of BB
		improves over SM for lower duty cycle as illustrated in the graph.  Here $\tau_{SM}$ and
		$\tau_{BB}$ indicate respectively SM and BB computational times for calculating 10-qubit propagators
		of $T = 0.5$ ms duration.}
	\label{bbscheme}
\end{figure}

Generally smooth modulations (SM) of the parameters $\{u_j(t)\}$ is assumed (see Fig. \ref{bbscheme}).
The overall quantum evolution is then evaluated by time discretization and subsequent calculation of instantaneous
propagators for each of the time steps.  
Here we propose a bang-bang (BB) control scheme which scales much more efficiently with the size of the control sequence.  Combined with a global optimization routine it can generate complex unitaries and even certain nonunitary operations with high fidelities. 

In the following we first briefly explain the BB scheme and then describe its applications in unitary and nonunitary controls.
We then illustrate the first experimental demonstration of optimal fixed-point quantum search (OFPQS) algorithm  on a three-qubit NMR quantum processor. Finally we also describe a nonunitary BB control for initializing an effective pure state in a five-qubit NMR system.

{\it Bang-bang approach:}
Although the following methods can be generalized to other quantum architectures, for the sake of clarity, we shall use the rotating frame picture of an NMR spin system.  
Denoting $\omega_r$, $J_{rs}$, $D_{rs}$ respectively for resonance off-sets, indirect (scalar), and direct coupling constants, and  $I_r$ for spin operators, the secular part of the internal Hamiltonian can be written in the form
\begin{eqnarray}
{\cal H}_\mathrm{int} = -\sum_r \omega_r I_{rz}
+ 2\pi \sum_{r<s} (J_{rs}+2D_{rs}) I_{rz}I_{sz} 
\nonumber \\
+ 2\pi \sum_{r<s}(J_{rs}-D_{rs})(I_{rx}I_{sx}+I_{ry}I_{sy}),
\end{eqnarray}  
wherein the third term vanishes for weakly interacting spins, particularly for spins belonging to different nuclear species
\cite{levittbook}.

The BB approach relies on intermittent bursts of full control power instead of its smooth modulation.
For generality we consider a spin-system having several nuclear species, and let the $j$th species  be irradiated by
an independent RF source with an amplitude switching between either $0$ or a maximum value $\Omega_j$, but
with a variable RF phase $\phi_j \in [0,2\pi]$ (see Fig. \ref{bbscheme}).  The propagator for the BB sequence
can be easily setup by discretizing the total control period $T$ into short segments of duration $\Delta t$.
Let ${\bf S}_{j} = \sum_m {\bf I}_{m}^{(j)}$ be the collective spin operator of $m$ spins belonging to the $j$th species.
We call $X_{j} = \exp\{-i ({\cal H}_0+ \Omega_{j} S_{jx}) \Delta t\}$ as the basic propagator and it needs to be computed
only once.  
During the $k$th segment there can be a delay or a pulse.
If the $k$th segment is a delay, then $U_{jk} = \exp(-i {\cal H}_0 \Delta t)$ is a constant operator and therefore
needs to be evaluated only once. 
On the other hand, if the $k$th segment is a pulse with a phase $\phi_{jk}$, then 
the propagator is simply
obtained by rotating $X_{j}$ about the $z$-axis, i.e., $U_{jk} = Z_{jk} X_j Z_{jk}^\dagger$.  Here
$Z_{jk} = \exp(-i\phi_{jk} S_{jz})$ is a diagonal operator in the Zeeman product basis and is therefore
efficiently evaluated in the run-time of the iterations.
The net propagator for the entire control sequence is simply the cumulative product
$U = \prod_{jk} U_{jk}$ over all the species $\{j\}$ and then over all the segments $\{k\}$.

Typically complex unitaries involving non-local quantum operations require long evolutions under spin-spin interactions, and have low duty cycles.  In such cases, the BB approach is orders of magnitude faster than the conventional methods which require repeated matrix exponentiations to evaluate the segment unitaries (Fig. \ref{bbscheme}).  If $[{\cal H}_0,Z_{jk}]=0$ (eg. weak-coupling case: $\vert \omega_r - \omega_s \vert \gg 2\pi \vert J_{rs} - D_{rs} \vert $ for all $r$ and $s$), then a further speed-up ensues, since the delay propagators become diagonal.

Given a target unitary $U_\mathrm{T}$, we optimize the BB parameters $\{\Omega_{jk},\phi_{jk}\}$ using the genetic algorithm and maximize the  unitary fidelity $F_u$.
If the goal is to prepare a specific quantum state $\rho_T$ starting from
an initial state $\rho_\mathrm{in}$, then we need to calculate the
output state $\rho_\mathrm{out} = U \rho_\mathrm{in} U^\dagger$ and the state fidelity
\begin{eqnarray}
F_s = \frac{\vert \mathrm{Tr} (\rho_T \rho_\mathrm{out})\vert}
{\sqrt{\mathrm{Tr}(\rho_T^2) \mathrm{Tr}(\rho_\mathrm{out}^2)}}
\end{eqnarray}
is to be maximized.  

In this work, we also describe preparing a target state with the help of the twirling operation which is essentially a nonunitary operator that attenuates all the coherences, ultimately retaining only the diagonal elements of the density matrix in the computational basis \cite{jones_twirling,abhishek_sspt}.  Although it is possible to incorporate the effects of decoherence, in our present scheme we ignore such effects by assuming that the control
sequences are much shorter than the decoherence time-scales.  In the following we briefly describe the application of BB sequence in implementing
a quantum algorithm.

{\it OFPQS algorithm:}
Classical search algorithms can find one or more `marked' items among an unsorted database of ${Q}$ items in $\mathcal{O}(Q)$ steps. On the other hand, Grover's quantum search algorithm achieves the same task in $\mathcal{O}(\sqrt{Q})$ steps, thereby providing a quadratic speedup over the classical counterpart \cite{originalgrover}. Grover's algorithm identifies one of the $R$ marked items among $Q$ unsorted items with the help of a given oracle function that can recognize the marked items. It can also be interpreted as a rotation in the 2D space spanned by the superposition of $Q- R$ non-solution states $\ket{\psi_{Q-R}}$ and the superposition of $R$ solution states $\ket{\psi_{R}}$ \cite{nielsenchuang}. The application of the Grover iterate can thus be visualized as the rotation of the initial state $\ket{\psi_{Q}}$ towards $\ket{\psi_{R}}$ in $\mathcal{O}(\sqrt{Q/R})$ steps. However, if we do not know the number of marked items $R$ beforehand, we cannot predict the number of iterations which would land the initial state closest to the marked state. Too few iterations give us a state comprising of mostly non-solution states, whereas, too many iterations can surpass the solution states and we may end up getting non-solution states, yet again.
In order to overcome this problem, attempts have been made to develop fixed point quantum search (FPQS) algorithms, which monotonically amplify the probability of obtaining the marked states \cite{groverfixed, tulsi_fixedpoint}.  While these FPQS algorithms lacked the quadratic speedup, a recent 
optimal FPQS  (OFPQS)  algorithm proposed by Yoder \textit{et. al.} achieves this speedup while maintaining the fixed point behavior
\cite{ofpqsa}.  In the following we outline the various steps involved in the OFPQS algorithm.

\begin{figure}
	\includegraphics[trim=0cm 9.35cm 0cm 2cm, clip=true,width=9cm]{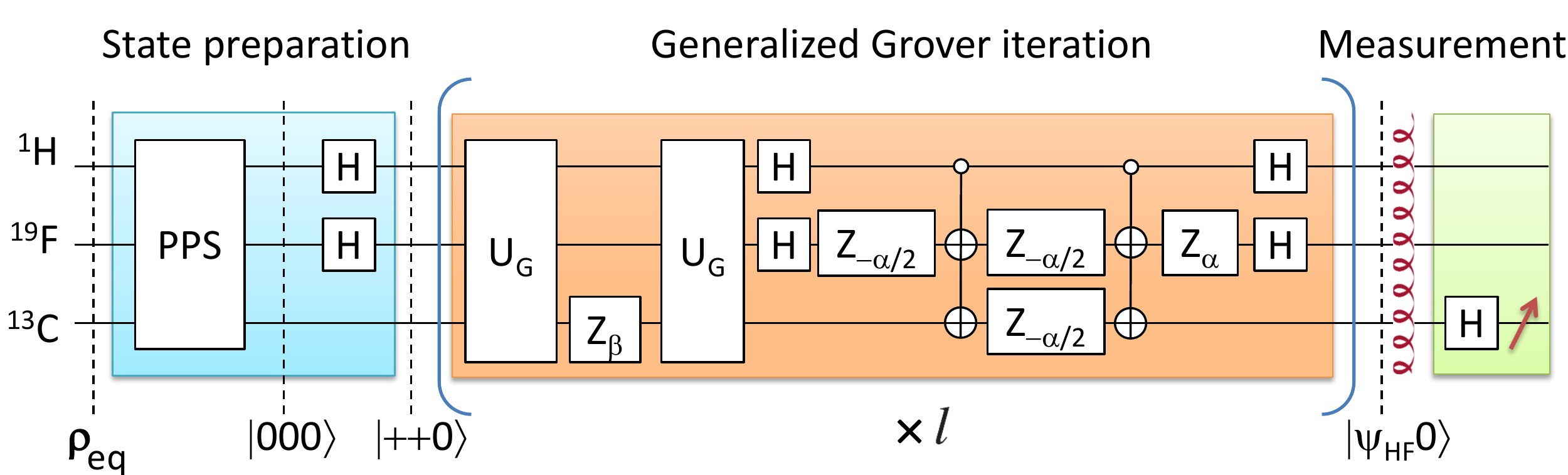}
	\caption{Quantum circuit for OFPQS algorithm.  }
	\label{circuit}
\end{figure}

The quantum circuit for the OFPQS algorithm is shown in Fig. \ref{circuit} (for 3-qubits).
We need to extract $\ket{\psi_{R}}$ from $\ket{\psi_{Q}}$ with a success probability $P_{L}$ with a predefined lower bound.  The algorithm needs a quantum register with a total number of qubits $n \ge \log_2(Q)+1$ including an ancilla qubit, all initialized in 
the ground state $\ket{0}^{\otimes n}$.
The system qubits are then transformed into a uniform superposition by applying $n-1$ Hadamard gates. 
We are provided with an oracle $U_G$ which when acted upon the marked state, flips the ancilla qubit. Thus, $U_G\ket{\psi_{R}}\ket{a}=\ket{\psi_{R}}\ket{a\oplus 1}$ and $U_G\ket{\psi_{Q-R}}\ket{a}=\ket{\psi_{Q-R}}\ket{a}$, 
where $\ket{a}$ represents the ancilla qubit.
Various gates in the generalized Grover iteration are also shown in Fig. \ref{circuit}.  
Let us consider $l$ generalized Grover iterations.  
Defining  $L=2l+1$ and $\gamma^{-1}=T_{1/L}(1/\delta)$ 
where $T_{L}(x)=\cos(L\cos^{-1}x)$ is the $L^{th}$ Chebyshev polynomial of the first kind, 
the phase rotations are given by
\begin{equation}
\alpha_{j}=-\beta_{l-j+1}=2\cot^{-1} \left(\tan(2\pi j/L)\sqrt{1-\gamma^{2}}\right) 
\end{equation}
for all $j = 1, 2, \dots l$ \cite{ofpqsa}.
 As each Grover iterate contains two applications of the oracle $U_G$, the query complexity is $L-1$. 

\begin{figure}
	\includegraphics[trim=2.8cm 0cm 5cm 0cm, clip=true,width=8cm]{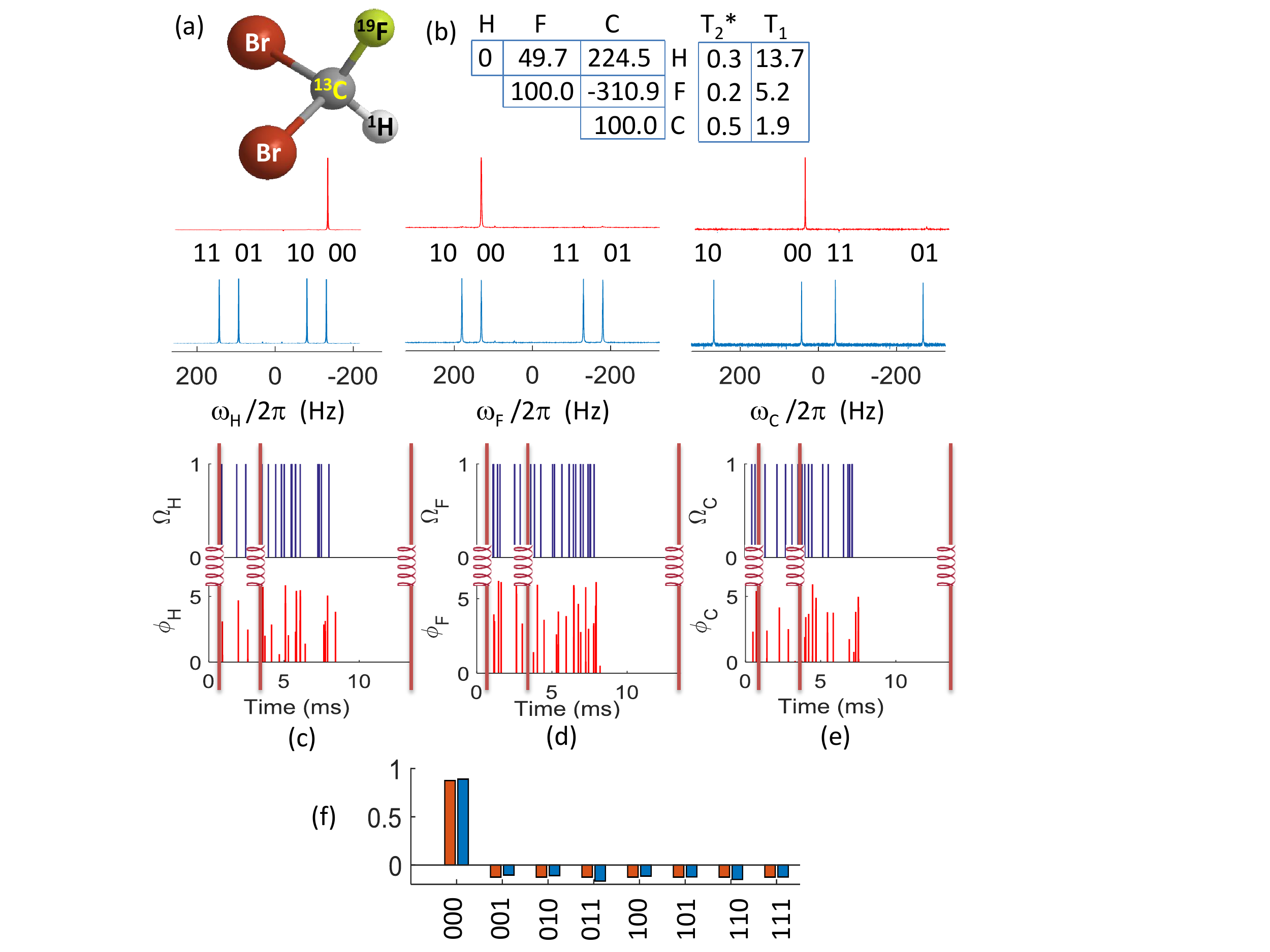} 
	\caption{(a) Molecular structure of dibromo fluoromethane, (b) the Hamiltonian parameters and relaxation time constants, and (c-e) 
		PPS spectra (upper trace), equilibrium spectra (middle trace), and PPS pulse-sequences (bottom trace) for the
		$^1$H, $^{19}$F and $^{13}$C qubits, (f) bar diagram representing theoretical (red) and experimental (blue) diagonal elements of the traceless deviation density matrix corresponding to $\ket{000}$ PPS. 
		In (b), the diagonal and off-diagonal elements are respectively chemical shifts
		and J-coupling constants in Hz.}
	\label{ppsfig}
\end{figure}

{\it NMR Implementation :}
We implement the  OFPQS algorithm on a quantum  register involving two system qubits $^1$H and $^{19}$F, along with an ancilla qubit $^{13}$C of dibromofluoromethane \cite{PhysRevLett.106.080401} dissolved in acetone-D6 (see Fig. \ref{ppsfig}). All the experiments were carried out on a Bruker 500 MHz NMR spectrometer at an ambient temperature of 300 K. 
Here, we  demonstrate the optimal search algorithm for searching one and two marked items among four unsorted items.  As described in the
circuit in Fig. \ref{circuit}, the experiment mainly involves three stages: (i) preparation of $\ket{000}$ state, (ii) iterations of the generalized Grover operation, and (iii) measurement of the final probabilities using the ancilla qubit. In the following each of the above stages is discussed in detail.

{\it Preparation of the initial state by nonunitary control:}
The equilibrium state of the 3-qubit register is represented by the density matrix $\rho_{eq} = (\mathbbm{1}+\epsilon_\mathrm{H} I_{z\mathrm{H}} + \epsilon_\mathrm{F} I_{z\mathrm{F}} + \epsilon_\mathrm{C} I_{z\mathrm{C}})/8$, where $\epsilon_i \sim 10^{-5}$ are called the purity factors. In NMR quantum information processing, one typically realizes a pseudopure state (PPS) that is isomorphic to a pure state using a combination of unitary and nonunitary processes \cite{corypps}.  
Fig. \ref{ppsfig} shows the BB-sequence consisting of both RF pulses and twirling operations automatically generated for the 3-qubit initialization by maximizing the state-to-state fidelity $F_s$. In  the experiment, the twirling process is easily achieved using three pulsed-field-gradients (PFGs)  whose time instants are obtained by BB optimization scheme.  Fidelity of the experimental results of  $\ket{000}$ PPS preparation shown in Fig. \ref{ppsfig} (c-e) is estimated to be $0.998 \pm 0.001$.  
Thus by assigning genes for the locations of the twirling operations along with other genes representing the RF pulses and
delays, one can realize a nonunitary operation.

As shown in the circuit of Fig. \ref{circuit}, we then prepare the uniform superposition state $\ket{\psi_Q}$ by applying a Hadamard gate on each of the system qubits.  

{\it Generalized Grover iteration by unitary control:}
The circuit for the generalized Grover iteration in Fig. \ref{circuit} is realized by a single BB sequence with an average fidelity of about 0.98 over 10\% RF inhomogeneity. The durations of the BB sequences for $l=\{1,2,\dots,10\}$ varied in the range 40 ms to 220 ms, and
consisted of 8,000 to 44,000 segments, each segment being $5\mu s$ long.  While such long sequences implementing multiple control operations 
are hard to realize by other control methods, the present BB scheme exploits the very low duty-cycle and the one-time evaluation of 
basis propagators to efficiently compute the overall unitaries.
We performed the experiments for one and two marked items by systematically increasing the number of iterations ($l$) in each case. 

\begin{figure}[b]
	\includegraphics[trim=0cm 0cm 0cm 0cm, clip=true,width=8.5cm]{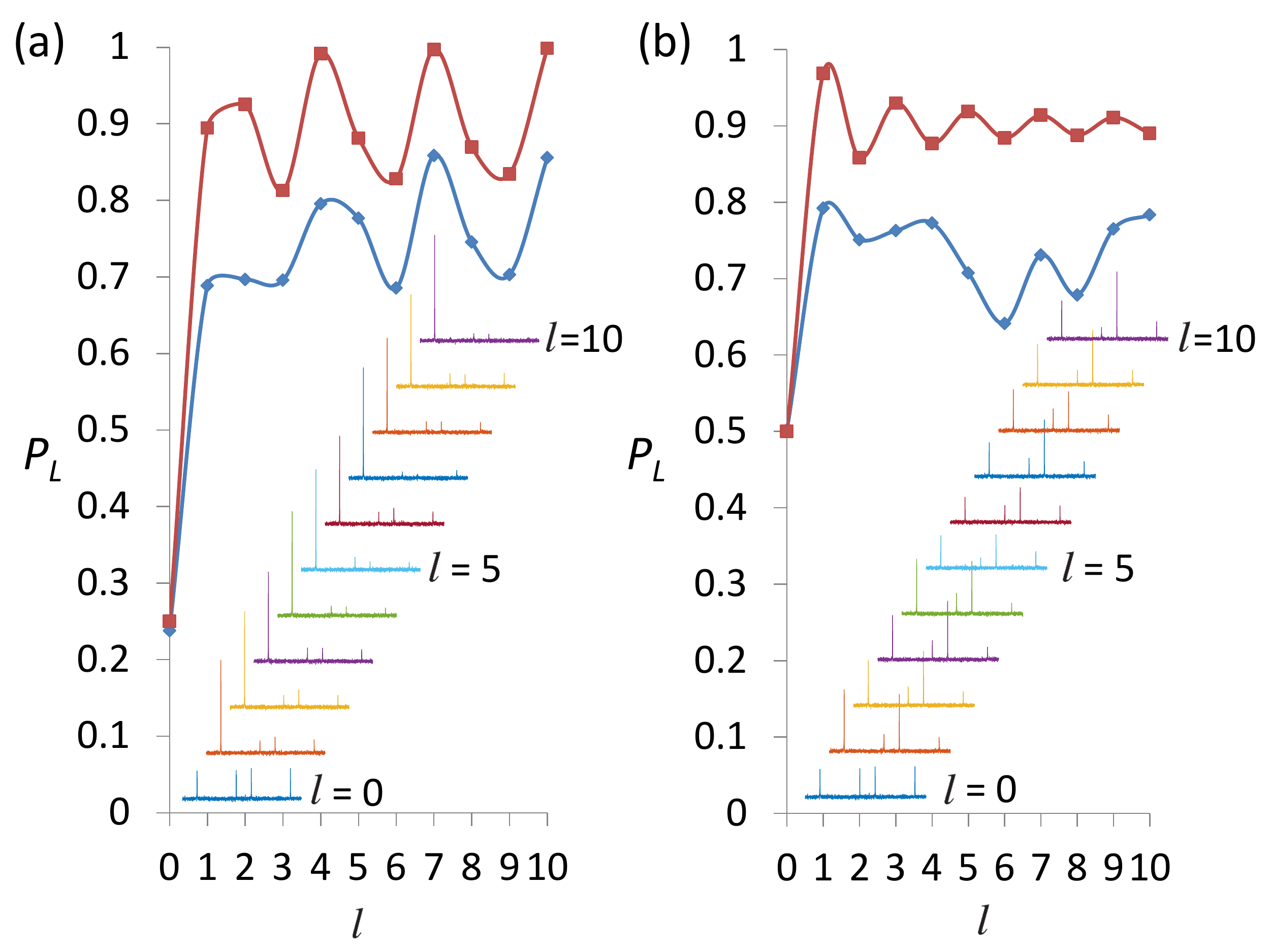}
	\caption{The theoretical and experimental results for the probability $P_L$ of finding  (a) one ($\ket{10}$) and (b) two
		($\ket{10}$ and $\ket{11}$) marked states among four items versus the number of iterations $l$. The red and blue points respectively represent the theoretically predicted and experimentally obtained probabilities measured directly from the ancilla ($^{13}$C) spectra (shown in insets).}
	\label{mark12}
\end{figure}

{\it Extracting the solutions by measuring ancilla:} In order to measure the final probabilities of finding the marked states, we first destroy the coherences by applying a twirling operation with the help of a strong PFG (Fig. \ref{circuit}).  The relative probabilities of the system qubits being in various eigenstates are encoded in the corresponding transitions of the ancilla qubit.  Therefore we finally measure the ancilla after applying a Hadamard operator, as shown in Fig. \ref{circuit}.

The experimental results of the probabilities versus iteration number for one and two marked states are shown in Fig. \ref{mark12}(a) and (b) respectively. While the theoretical lower bounds for the probabilities of finding the marked states are 0.8, the experimental lower bounds were 0.69 and 0.64 for one and two marked states respectively.  The lower values of experimental probabilities are mainly due to pulse imperfections and decoherence.  In spite of the lower probabilities, the marked states can be clearly identified from the spectra.
It is clear that as the number of iterations increases, each final state remains close to the solution state, thus exhibiting the fixed point behavior.

\begin{figure}[b]
	\includegraphics[trim=0cm 1.0cm 0cm 0cm, clip=true,width=9cm]{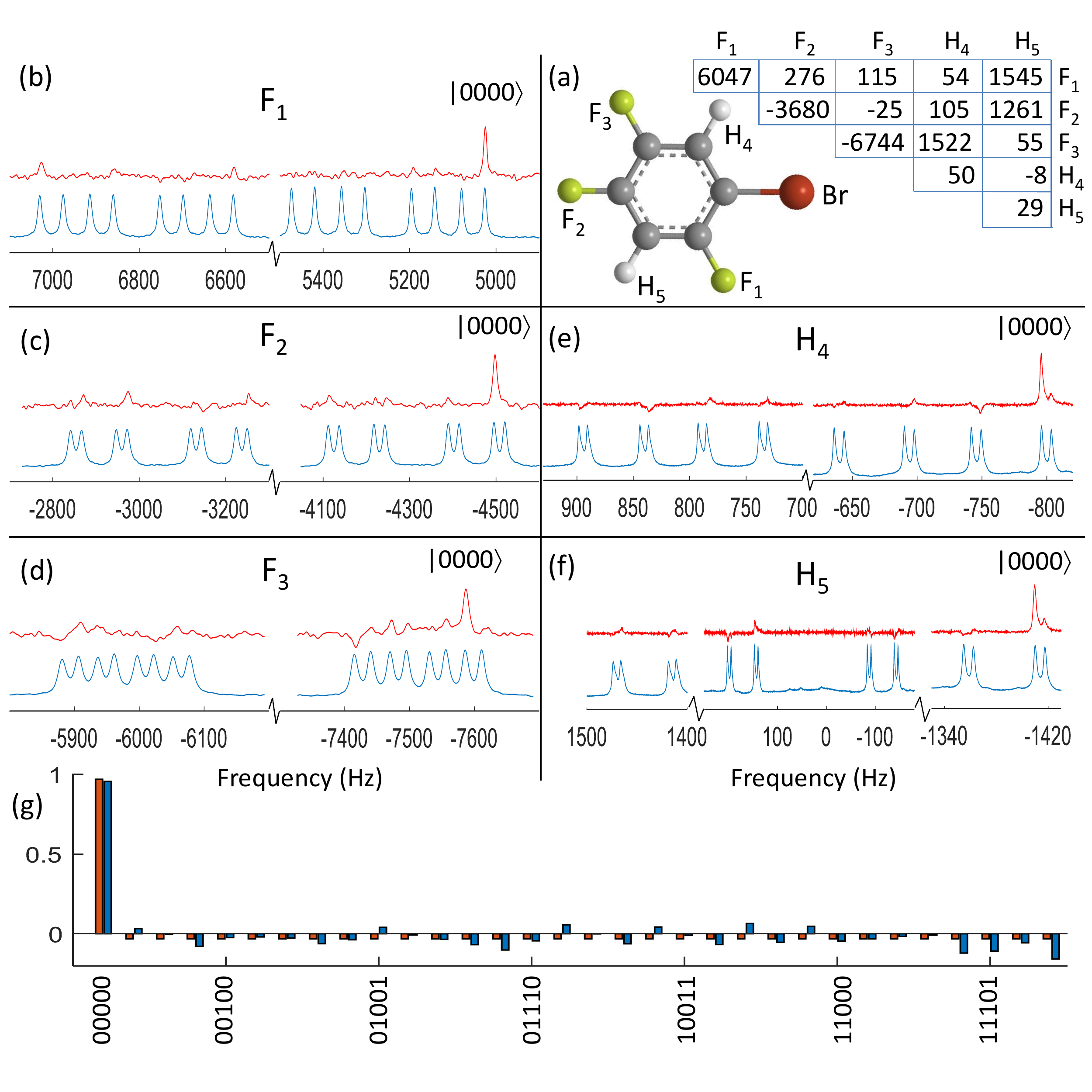}
	\caption{(a) Molecular structure of the five-qubit system 1-bromo-2,4,5-trifluorobenzene and its
		Hamiltonian parameters wherein diagonal and off-diagonal numbers represent chemical shifts
		and J-couplings (in Hz), (b-f) the spectra corresponding to the thermal equilibrium state (blue)
		and $\ket{00000}$ PPS prepared from the nonunitary BB sequence (red), and (g) 
		bar diagram representing theoretical (red) and experimental (blue) diagonal elements of the traceless 
		deviation density matrix corresponding to $\ket{00000}$ PPS.} 
	\label{btfbz}
\end{figure}

{\it Preparation of PPS in a 5-qubit system:}
To demonstrate the potential of the BB scheme in a larger quantum register, we 
carried out preparation of pseudopure state in a five-qubit system consisting
of three $^{19}$F and two ${^1}$H spins of 1-bromo-2,4,5-trifluorobenzene
(see inset of Fig. \ref{btfbz}) partially oriented in a liquid crystal 
\cite{abhishek_aaqst}.  The nonunitary transformation was realized using a
single BB sequence of 89.1 ms duration involving four twirling operations.
The normal and PPS spectra of each spin are also shown in Fig. \ref{btfbz}.  
The overall fidelity of the PPS preparation was estimated to be $0.96 \pm 0.02$.  The lower
fidelity compared to the 3-qubit register is not only due to the increased
complexity, but also due to the temporal fluctuations in the dipolar 
interaction strengths of the 5-qubit system.  Nevertheless, the automatic procedure
for generating the nonunitary transformation promises applications in a variety of 
physical implementations involving quantum control.

{\it Conclusions:}
Bang-bang pulses are efficient to compute, easier to implement, and are 
robust against pulse errors.  
Using a nonlocal optimization algorithm, such
as the genetic algorithm, it is possible to efficiently optimize the BB sequence
to generate any desired unitary transformation or to prepare a quantum state.
For example, in the case of NMR, long 
control sequences are often needed to apply nonlocal transformations.
It is unnecessary to apply RF pulses throughout the control sequence,
which is not only hard to compute but also prone to RF inhomogeneity errors.
The BB control technique allowed us to carry-out the first experimental demonstration of
optimal fixed point quantum search algorithm on a three-qubit NMR register.
The experimental spectra easily identified the marked states out of a database
of four unsorted items.  The entire generalized Grover iterate, encompassing
several local and nonlocal gates, has been realized using a single BB-sequence 
lasting more than 200 ms, involving over 40,000 segments.  
Such long control sequences are hard to compute by other optimal control techniques.
Combined with twirling operations it is also possible to realize nonunitary transformations.  
Using this scheme, we have demonstrated the synthesis 
and experimental implementation of pseudopure states in 3- as well as 5-qubit NMR quantum 
registers. 
BB sequences can be realized by on and off switching of the control field and therefore
it is also applicable to a variety of other architectures where smooth amplitude modulation
of control fields is difficult.
It may also be possible to realize a hybrid control sequence by combining the
BB scheme with other optimal control techniques.  Although the present experimental demonstration
is in an NMR system, the simplicity of the BB scheme allows its application in other
architectures such as SQUID or NV$^-$ center-based quantum registers.

\textit{Acknowledgments:}
We acknowledge useful discussions with Dr. Abhishek Shukla, Swathi Hegde, Sudheer Kumar,
and Deepak Khurana.  We also thank Prof. Anil Kumar, IISc, Bangalore for providing us the three-qubit
NMR register.

\bibliographystyle{apsrev4-1}
\bibliography{bibfq}

\end{document}